\documentclass[preprint,showpacs,preprintnumbers,amsmath,amssymb]{revtex4}
\usepackage{graphicx}
\usepackage{dcolumn}
\usepackage{bm}
\usepackage{color}
\usepackage{hhline}

\begin{document}

\title{Theoretical study of the role of the tip in enhancing the sensitivity of differential conductance tunneling spectroscopy
       on magnetic surfaces}

\author{Kriszti\'an Palot\'as}
\email{palotas@phy.bme.hu}
\affiliation{Budapest University of Technology and Economics,
Department of Theoretical Physics, Budafoki \'ut 8., H-1111 Budapest, Hungary}

\author{Werner A. Hofer}
\affiliation{University of Liverpool, Surface Science Research Centre, L69 3BX Liverpool, UK}

\author{L\'aszl\'o Szunyogh}
\affiliation{Budapest University of Technology and Economics,
Department of Theoretical Physics, Budafoki \'ut 8., H-1111 Budapest, Hungary }

\date{\today}

\begin{abstract}

Based on a simple model for spin-polarized scanning tunneling spectroscopy (SP-STS) we study how tip magnetization and
electronic structure affects the differential conductance ($dI/dV$) tunneling spectrum of an Fe(001) surface.
We take into account energy dependence of the vacuum decay of electron states, and tip electronic structure
either using an ideal model or based on ab initio electronic structure calculation.
In the STS approach, topographic and magnetic contributions to $dI/dV$ can clearly be distinguished and analyzed separately.
Our results suggest that the sensitivity of STS on a magnetic sample can be tuned and even enhanced by choosing the appropriate
magnetic tip and bias set point, and the effect is governed by the effective spin-polarization.

\end{abstract}


\maketitle

\section{Introduction}

Development of simulation tools for scanning tunneling microscopy (STM) and spectroscopy (STS) are in the focus of theorists
\cite{hofer03rmp,hofer03pssci} since the invention of STM 30 years ago.
It has been established that the tip electronic structure plays a crucial role in measured differential conductance ($dI/dV$)
tunneling spectra, see e.g.\ Refs.\ \cite{passoni09,kwapinski10}.
While a theoretical method has been proposed to separate tip and sample contributions to STS \cite{hofer05sts},
some recent research activities have focused on extracting surface local electronic properties from experimental STS data
\cite{ukraintsev96,koslowski07,passoni09,ziegler09,koslowski09},
which is the convolution of tip and sample electronic structures.
The situation is expected to be even more complicated in magnetic systems due to effective spin-polarization.

STM can be made sensitive to magnetism, and spin-polarized scanning tunneling microscopy (SP-STM) and spectroscopy (SP-STS)
are nowadays the key tools for nanomagnetic research \cite{bode03review,wiesendanger09review,wulfhekel10review},
i.e.\ for studying and manipulating magnetic properties of surfaces and deposited magnetic nanoclusters with atomic scale
resolution \cite{weiss05,tao09,serrate10}.
SP-STS has recently been used to find inversion of spin-polarization above magnetic adatoms \cite{yayon07,heinrich09,zhou10},
which effect has been explained theoretically \cite{ferriani10tip}. This spectroscopic approach turned out to be useful for
investigating many-body effects on substrate-supported adatoms \cite{neel07kondo,ternes09sts} as well.
Tip effects on SP-STS \cite{rodary09} and on achieving giant magnetic contrast \cite{hofer08tipH} have also been reported.
Recent experiments of Schouteden et al.\ \cite{schouteden08} and Heinrich et al.\ \cite{heinrich10} show evidence that
STS peaks are different above magnetic islands of opposite magnetization. However, no detailed explanation has been given
for these observations. In our present study we shed light on this effect in a different setup by studying tunneling spectra
of the magnetic Fe(001) surface. Our results show the importance of the effective spin-polarization on the STS spectra,
which can be tuned by changing tip magnetization direction or bias set point for the tip.

The paper is organized as follows: Theoretical model of SP-STS is presented in section \ref{sec_spsts}.
Based on this, simulating differential tunneling spectra of the Fe(001) surface is presented and discussed in
section \ref{sec_res}. We focus particularly on tip effects, and consider an ideal electronically flat and maximally
spin-polarized tip as well as a more realistic ferromagnetic Ni tip.
Our results suggest that the sensitivity of tunneling spectroscopy measurements on magnetic surfaces can be enhanced by using
the proper magnetic tip and bias set point. Summary of our findings is found in section \ref{sec_conc}.

\section{Theoretical model of SP-STS}
\label{sec_spsts}

Based on the Transfer Hamiltonian approach, the work of Passoni and Bottani \cite{passoni07} reports an advanced way to
incorporate results of electronic structure calculations into STS simulations.
Our model for SP-STS is based on the spin-polarized version of the Tersoff-Hamann model \cite{tersoff83,tersoff85}
introduced by Wortmann et al.\ \cite{wortmann01}, which is adapted within the framework of the atom-superposition approach
\cite{tersoff85,yang02,smith04,heinze06}. Approximations and limitations of this approach have been discussed in the literature
\cite{tersoff85,smith04}.
By simulating differential conductance, we directly calculate the differential spectrum and not the numerical derivative of an
integral spectrum, see Ref.\ \cite{hofer05sts}. This way we assume that $dI/dV$ is proportional to the electron local density of
states ($LDOS$), which is a reasonable approximation at low bias voltages.
Calculating $dI/dV$ from the tunneling current by numerical differentiation is a commonly used approach
\cite{passoni09,koslowski07,ziegler09,passoni07}. The motivation comes from experiments, where the surface $LDOS$ is not known,
and it is the goal to extract this quantity from measurements. In our SP-STS approach based on first principles electronic
structure calculations we proceed in the opposite way: We calculate surface electronic structure, define $LDOS$ at tip apex
position, also taking into account tip electronic structure, and then we define $dI/dV$ proportional to $LDOS$.
This way, the so-called background term does not explicitly occur in our $dI/dV$ expression, which is an important correction to
$dI/dV$ at higher bias voltages, see Ref.\ \cite{passoni09}, Eq.(3), second term. Assuming an electronically featureless tip,
this term is approximated to be proportional to the tunneling current \cite{koslowski07,ziegler09}, and, thus, it is expected
to increase with increasing absolute value of the bias voltage.
Due to the energetic position of the Fe(001) surface state peak close to the Fermi level, in the present work we neglect the
background term and we will deal with its relevance in the future. We expect, however, that the inclusion of the background
term does not affect our conclusions for tuning the $dI/dV$ peaks by using a magnetic tip as long as staying in the low bias
regime. Resolving $dI/dV$ features at higher bias voltages turned out to be difficult even for using nonmagnetic tips and,
therefore, different normalization schemes have been introduced to obtain information about the sample local electronic structure
\cite{ukraintsev96,koslowski07,passoni09,ziegler09,koslowski09}.

Incorporating bias dependence into our model one has to take into account energy dependence of the vacuum decay of electron states.
Moreover, we show how to explicitly incorporate energy dependence of tip electronic structure based on the result from
ab initio calculation into our bias dependent atom-superposition-based model.
This means that different tip models and their effect on tunneling properties can be investigated.
The only requirement for our present formalism is that we assume that electrons tunnel through one tip apex atom.

In order to simulate single point differential tunneling spectra above the surface atom with lateral coordinates $(x_0,y_0)$,
first, we calculate the $LDOS$ at the tip apex position $\underline{R}_{TIP}(x_0,y_0,z)$.
The spin-mixed $LDOS$ at a given energy $E$ can be decomposed into a non-spin-polarized ($TOPO$), and a spin-polarized ($MAGN$)
part as
\begin{equation}
\label{Eq_LDOS_decomp}
LDOS(x_0,y_0,z,E)=LDOS_{TOPO}(x_0,y_0,z,E)+LDOS_{MAGN}(x_0,y_0,z,E).
\end{equation}
Using the atom-superposition method \cite{smith04,heinze06}, the two terms can be written as
\begin{eqnarray}
\label{Eq_LDOSTOPO}
LDOS_{TOPO}(x_0,y_0,z,E)&=&\Delta E\sum_{\alpha}e^{-2\kappa(E)\left|\underline{R}_{TIP}(x_0,y_0,z)-\underline{R}_{\alpha}\right|}n_T(E)n_S^{\alpha}(E)\\
LDOS_{MAGN}(x_0,y_0,z,E)&=&\Delta E\sum_{\alpha}e^{-2\kappa(E)\left|\underline{R}_{TIP}(x_0,y_0,z)-\underline{R}_{\alpha}\right|}m_T(E)m_S^{\alpha}(E)cos\varphi_{\alpha},
\label{Eq_LDOSMAGN_coll}
\end{eqnarray}
where the sum over $\alpha$ has to be carried out over all the surface atoms with position vectors $\underline{R}_{\alpha}$,
each characterized by a local spin quantization axis determined by their atomic spin moment direction.
$\Delta E$ is the energy resolution for our simulated tunneling spectra, and it ensures that the $LDOS$ is correctly
measured in units of $1/eV$. A $\Delta E$ value of $10^{-3}$ $eV$ has been used in our calculations.
The exponential factor is the transmission coefficient for electrons tunneling between states of atom $\alpha$ on the surface and
the tip apex, where $\kappa$ is the vacuum decay. $\kappa$ is treated within the independent-orbital approximation
\cite{tersoff83,tersoff85,heinze06}, which means that the same decay is used for all type of orbitals,
but its energy dependence is explicitly considered in the same fashion as in Ref.\ \cite{lang86}.
Extension of our model in the direction to incorporate orbital dependent vacuum decay following Chen's work \cite{chen90}
is planned in the future. In the present paper we propose two different ways of calculating
$\kappa$, one is inspired by the Tersoff-Hamann model, taking only surface properties into account,
\begin{equation}
\label{Eq_kappa_TH}
\kappa(E)=\frac{1}{\hbar}\sqrt{2m(\phi_S+E_F^S-E)},
\end{equation}
where the electron's mass is $m$ and charge $-e$, while
$\phi_S$ and $E_F^S$ are the average electron workfunction and the Fermi energy of the sample surface, respectively.
We use this energy dependent vacuum decay for an ideal, electronically featureless and maximally spin-polarized tip model.
The second expression for $\kappa$ is based on the Wentzel-Kramers-Brillouin (WKB) approximation assuming a rectangular tunnel
barrier,
\begin{equation}
\kappa(E,V)=\frac{1}{\hbar}\sqrt{2m\left(\frac{\phi_S+\phi_T+eV}{2}+E_F^S-E\right)},
\label{Eq_kappa_WKB}
\end{equation}
with $\phi_T$ being the local electron workfunction of the tip apex, and $V$ is the applied bias voltage.
This vacuum decay formula is considered for our magnetic Ni tip model. The quantity $(\phi_S+\phi_T+eV)/2+E_F^S-E$
is the energy- and bias dependent apparent barrier height for tunneling electrons, $\phi_a(E,V)$.
Note that our vacuum decay formulae are asymmetric with respect to positive and negative bias regime \cite{ukraintsev96}.
The average workfunction of the sample surface is calculated from the local electrostatic potential on a
three-dimensional fine grid, $V(x,y,z)$, as
\begin{equation}
\label{Eq_WF_average}
\phi_S=\max_z\left\{\frac{1}{N_xN_y}\sum\limits_{x,y}V(x,y,z)\right\}-E_F^S,
\end{equation}
with $N_x$ and $N_y$ the corresponding number of grid points,
and the local workfunction of tip apex is obtained as
\begin{equation}
\label{Eq_WF_local}
\phi_T=\max_z\left\{V(x_0,y_0,z)\right\}-E_F^{TIP},
\end{equation}
with $x_0$ and $y_0$ lateral coordinates of the tip apex atom, and $E_F^{TIP}$ the Fermi energy of the tip material.

In the above $LDOS$ formulae, $n_T(E)$ and $n_S^{\alpha}(E)$ denote electron charge DOS projected to the tip apex and the
$\alpha$th surface atom, respectively,
\begin{eqnarray}
n_T(E)&=&n_T^{\uparrow}(E)+n_T^{\downarrow}(E),\nonumber\\
n_S^{\alpha}(E)&=&n_S^{\alpha\uparrow}(E)+n_S^{\alpha\downarrow}(E),
\label{Eq_CDOS_coll}
\end{eqnarray}
$\uparrow$ and $\downarrow$ relative to their local spin quantization axes.
Similarly, $m_T(E)$ and $m_S^{\alpha}(E)$ are electron magnetization DOS (MDOS) projected to the tip apex and the
$\alpha$th surface atom, respectively,
\begin{eqnarray}
m_{T}(E)&=&n_{T}^{\uparrow}(E)-n_{T}^{\downarrow}(E),\nonumber\\
m_S^{\alpha}(E)&=&n_S^{\alpha\uparrow}(E)-n_S^{\alpha\downarrow}(E).
\label{Eq_MDOS_coll}
\end{eqnarray}
$\varphi_{\alpha}$ is the angle between the spin moments of the tip apex and the $\alpha$th surface atom. Above, the
spin-resolved atom-projected DOS ($PDOS$) quantities, $n_T^{\uparrow,\downarrow}(E)$ and $n_S^{\alpha\uparrow,\downarrow}(E)$,
are calculated from first principles. For this task any available ab initio electronic structure code can be used.
Spin-resolved $PDOS$ can also be calculated at finite temperatures, if we assume a Gaussian broadening of the peaks at the
k-resolved spin-dependent electron energy (Kohn-Sham) eigenvalues, $\varepsilon_{T,S}^{j\uparrow,\downarrow}(\underline{k})$,
obtained at zero temperature, as
\begin{eqnarray}
n_T^{\uparrow,\downarrow}(E)&=&\sum_{\underline{k}}\sum_j\frac{1}{G\sqrt{\pi}}e^{-\left(E-\varepsilon_T^{j\uparrow,\downarrow}(\underline{k})\right)^2/G^2}\int\limits_{V_{\{tip\;apex\;atom\}}}d^3r\Psi_T^{j\underline{k}\uparrow,\downarrow\dagger}(\underline{r})\Psi_T^{j\underline{k}\uparrow,\downarrow}(\underline{r}),\nonumber\\
n_S^{\alpha\uparrow,\downarrow}(E)&=&\sum_{\underline{k}}\sum_j\frac{1}{G\sqrt{\pi}}e^{-\left(E-\varepsilon_S^{j\uparrow,\downarrow}(\underline{k})\right)^2/G^2}\int\limits_{V_{\{\alpha th\;surface\;atom\}}}d^3r\Psi_S^{j\underline{k}\uparrow,\downarrow\dagger}(\underline{r})\Psi_S^{j\underline{k}\uparrow,\downarrow}(\underline{r}),
\label{Eq_PDOS}
\end{eqnarray}
with $\Psi_{T,S}^{j\underline{k}\uparrow,\downarrow}(\underline{r})$ the spin-dependent electron wavefunctions corresponding to
$\varepsilon_{T,S}^{j\uparrow,\downarrow}(\underline{k})$ for tip ($T$) and surface ($S$), respectively,
and $j$ the energy band index.
The integral over the atomic volumes can be performed either in the atomic sphere or within the Bader volume \cite{tang09}.
In the present study we use integral over atomic spheres.
The Gaussian parameter $G$ could, in general, be temperature dependent. In our slab calculations, we fixed its value to
0.1 $eV$, which always provided smooth $n_T^{\uparrow,\downarrow}(E)$ and $n_S^{\alpha\uparrow,\downarrow}(E)$ functions.

$LDOS$ can also be written in terms of energy dependent spin-polarizations, $P(E)$, as
\begin{equation}
\label{Eq_LDOS_Spinpol}
LDOS(x_0,y_0,z,E,V)=\Delta E\sum_{\alpha}e^{-2\kappa(E,V)\left|\underline{R}_{TIP}(x_0,y_0,z)-\underline{R}_{\alpha}\right|}n_T(E)n_S^{\alpha}(E)[1+P_T(E)P_S^{\alpha}(E)cos\varphi_{\alpha}],
\end{equation}
where the spin-polarization is defined as
\begin{eqnarray}
P_T(E)&=&\frac{m_T(E)}{n_T(E)}=\frac{n_T^{\uparrow}(E)-n_T^{\downarrow}(E)}{n_T^{\uparrow}(E)+n_T^{\downarrow}(E)},\nonumber\\
P_S^{\alpha}(E)&=&\frac{m_S^{\alpha}(E)}{n_S^{\alpha}(E)}=\frac{n_S^{\alpha\uparrow}(E)-n_S^{\alpha\downarrow}(E)}{n_S^{\alpha\uparrow}(E)+n_S^{\alpha\downarrow}(E)}.
\label{Eq_Spinpol_coll}
\end{eqnarray}
Using Eq.(11) of Ref.\ \cite{wortmann01} and our $LDOS$ expression, the differential conductance
at the tip apex position and at energy $E$ is
\begin{equation}
\label{Eq_dIdV}
\frac{dI}{dV}(x_0,y_0,z,E,V)=\frac{e^2}{h}(\Delta E)^2\sum_{\alpha}e^{-2\kappa(E,V)\left|\underline{R}_{TIP}(x_0,y_0,z)-\underline{R}_{\alpha}\right|}n_T(E)n_S^{\alpha}(E)[1+P_T(E)P_S^{\alpha}(E)cos\varphi_{\alpha}].
\end{equation}
This is an important step to define $dI/dV$ without the need for calculating the tunneling current.
Multiplying the $LDOS$ with $\Delta E$ results in a dimensionless quantity, which is multiplied by the conductance quantum $e^2/h$
in order to arrive at our $dI/dV$ expression.
This means that $n_T(E)\Delta E$ electron states from tip and $n_S^{\alpha}(E)\Delta E$ states from each surface atom
contribute to the differential conductance at energy $E$, and in our model, $dI/dV$ is proportional to the $LDOS$, which contains
both surface and tip electronic information. Electronic structure ($PDOS$) of tip apex can be calculated at the same level as
surface electronic properties, but it is also possible to combine different levels of electronic structure calculations, or
include simplified model tip electronic structures into our approach.
For example, assuming an electronically flat maximally spin-polarized ($P_T(E)=1$) ideal magnetic tip with
e.g.\ $n_T(E)=1/\Delta E$, the differential conductance reads
\begin{equation}
\label{Eq_dIdV-idealtip}
\frac{dI}{dV}(x_0,y_0,z,E)=\frac{e^2}{h}\Delta E\sum_{\alpha}e^{-2\kappa(E)\left|\underline{R}_{TIP}(x_0,y_0,z)-\underline{R}_{\alpha}\right|}n_S^{\alpha}(E)[1+P_S^{\alpha}(E)cos\varphi_{\alpha}].
\end{equation}
Here, Eq.(\ref{Eq_kappa_TH}) has been assumed for the vacuum decay, and there is no $V$ dependence.

In single point STS experiments the tip is fixed above a surface atom. Its $z$ position is determined by the bias set point
$V$ and the tunneling current $I$. Additionally, a modulating voltage $V_M$ with a small amplitude is added to $V$, and $dI/dV$
can be obtained from measuring current modulation \cite{wiesendanger09review}. Our model corresponds to this experimental setup,
and the bias set point $V$ determines the relative energetic position of sample and tip electronic structure, such that
$E_F^{TIP}=E_F^T=E_F^S+eV$, see Figure \ref{Fig1} for a sketch.
The tip set point can be given either by $V$ and $I$ or by $V$ and $z$ because there is a one to one correspondence
between $z$ and $I$ at given bias voltage, $V$, i.e.\ $z(x,y)$ corresponds to the height profile of
a constant current contour, $I=I_C$.
Note that in all of our calculated single point $dI/dV$ spectra we fix the tip apex $z$=3.5 $\AA$ above the surface Fe atom.

The general energy dependence of $dI/dV$ in Eq.(\ref{Eq_dIdV}) can be cast into bias dependence, $U$, which is varied,
while $V$ is fixed. Using $E=E_F^S+eU$ we obtain
\begin{eqnarray}
\label{Eq_dIdV-V}
\frac{dI}{dV}(x_0,y_0,z,U,V)&=&\frac{e^2}{h}(\Delta E)^2\sum_{\alpha}e^{-2\kappa(E_F^S+eU,V)\left|\underline{R}_{TIP}(x_0,y_0,z)-\underline{R}_{\alpha}\right|}\\
&\times&n_T(E_F^T+eU-eV)n_S^{\alpha}(E_F^S+eU)[1+P_T(E_F^T+eU-eV)P_S^{\alpha}(E_F^S+eU)cos\varphi_{\alpha}].\nonumber
\end{eqnarray}
Note that by integrating our $dI/dV$ expression with respect to $eU$ in a given energy window defined by the bias voltage $V$ and
temperature, the tunneling current can be calculated, which expression is identical to the starting point of other STS theories
\cite{passoni09,koslowski07,ziegler09,passoni07}, except the fact that our tunneling current is formulated within the
atom-superposition approach. This will be reported in the future. In the present study, we focus on the simulation of SP-STS.
Similarly to Eq.(\ref{Eq_LDOS_decomp}), the differential conductance can also naturally be decomposed into a
non-spin-polarized ($TOPO$) and a spin-polarized ($MAGN$) part,
\begin{equation}
\label{Eq_dIdV_decomp}
\frac{dI}{dV}(x_0,y_0,z,U,V)=\frac{dI_{TOTAL}}{dV}(x_0,y_0,z,U,V)=\frac{dI_{TOPO}}{dV}(x_0,y_0,z,U,V)+\frac{dI_{MAGN}}{dV}(x_0,y_0,z,U,V),
\end{equation}
thus, the contributions can be analyzed separately. This is also valid if we start from the tunneling current and
define the $dI/dV$ as its derivative with respect to bias voltage, see e.g.\ Ref.\ \cite{passoni07}, section III.A,
where the occurring extra background term can be straightforwardly separated into $TOPO$ and $MAGN$ parts containing
$\int dE n_S^{\alpha}(E)\frac{\partial}{\partial V}\left[e^{-2\kappa(E,V)\left|\underline{R}_{TIP}(x_0,y_0,z)-\underline{R}_{\alpha}\right|}n_T(E-eV)\right]$ and $cos\varphi_{\alpha}\int dE m_S^{\alpha}(E)\frac{\partial}{\partial V}\left[e^{-2\kappa(E,V)\left|\underline{R}_{TIP}(x_0,y_0,z)-\underline{R}_{\alpha}\right|}m_T(E-eV)\right]$ terms, respectively,
inherent to our atom-superposition approach.

The presented method for simulating STS can also be applied for nonmagnetic systems, where all magnetic contributions are equal to
zero.
Moreover, note that this method can be generalized in order to obtain two-dimensional differential conductance maps, however,
without knowing the tunneling current, only in the so-called constant distance mode, where the tip apex position is varied in a
plane parallel to the surface. This does not correspond to the usual experimental setup measuring tunneling spectra on a constant
current contour, which ensures constant tip-sample distance even for rough surfaces \cite{wiesendanger09review,ziegler09}.
The extension of our model in this direction is under way, and will be reported in a subsequent paper.

\section{Results and Discussion}
\label{sec_res}

We simulate differential conductance tunneling spectra of the Fe(001)
surface, measured with an ideal, electronically flat, maximally spin-polarized tip, and with a model ferromagnetic Ni tip.
The ideal tip has two advantages: (1) due to its electronically featureless character the bias set point does not play a role
and the sole effect of tip magnetization direction on the spectra can be investigated, and (2) assuming maximal
spin-polarization for the tip, the biggest magnetic effect on the spectra is expected.
On the other hand, electronic structure of the Ni tip apex atom has been calculated from first principles.

Tunneling spectrum of Fe(001) has been studied experimentally earlier using a nonmagnetic W tip \cite{stroscio95}. According to
that measurement the obtained peak at $+0.17$ V corresponds to a surface state of Fe(001) close to the $\Gamma$ point.
Ref.\ \cite{biedermann96} reports a slightly different peak position of the same surface state at $+0.3$ V.
This difference might be due to different tip geometry \cite{hofer00Fe}, however, to the best of our knowledge, so far
magnetic tips have not been considered for studying the Fe(001) surface.
Here, we demonstrate that this surface state peak can be tuned either by changing tip magnetization direction or bias set point.
The results suggest that the sensitivity of STS depend crucially on the tip and it can be enhanced by finding
favorable combination of these factors.

Spin-polarized collinear electronic structure calculations have been performed with standard Density Functional Theory (DFT)
methods within the Generalized Gradient Approximation (GGA) implemented in the Vienna Ab-initio Simulation Package (VASP)
\cite{VASP2,VASP3,hafner08}. A plane wave basis set for electronic wavefunction expansion together with the
projector augmented wave (PAW) method \cite{kresse99} has been applied, while
the exchange-correlation functional is parametrized according to Perdew and Wang (PW91) \cite{pw91}.

We calculated a 9-layer Fe slab, where the surface layer on one side and the first subsurface layer have been fully relaxed.
After relaxation the interlayer distances are reduced by 1.2\% and 6.2\% compared to bulk, respectively.
We also checked a thicker slab of 13-layer Fe, where both surface layers have been relaxed and found
no significant difference in the tunneling spectra. A separating vacuum region of 10 $\AA$ width in the
surface normal ($z$) direction has been set up between neighboring supercell slabs.
For calculating the projected electron DOS onto the surface iron atom in our $1\times 1$ surface unit cell
we used an $11\times 11\times 1$ Monkhorst-Pack (MP) \cite{monkhorst} k-point grid.
The magnetic moment of the Fe surface atom is 2.84 $\mu_B$.
The average electron workfunction of the iron surface is calculated to be $\phi_S=3.99$ $eV$ using Eq.(\ref{Eq_WF_average}).
The spin quantization axis of Fe points along the direction of its in-plane magnetic moment.

Figure \ref{Fig2} shows simulated differential tunneling spectra (solid lines) $z$=3.5 $\AA$ above the surface Fe atom
in the bias range from $-1.0$ V to $+1.0$ V using the ideal tip, employing Eq.(\ref{Eq_dIdV-idealtip}).
The spectra are rescaled such that they can be shown together with the spin-polarization, $P_S(E)$.
The total tunneling spectra (TOTAL) are decomposed into topographic ($TOPO$) and magnetic ($MAGN$) contributions,
according to Eq.(\ref{Eq_dIdV_decomp}). Focusing on the nonmagnetic contribution (TOPO, solid red curve),
one can immediately observe that we obtain STS peaks at $+0.2$ V and slightly below $+0.6$ V, where the former one is in
excellent agreement with previous experiment \cite{stroscio95}, and both peaks are in good agreement with
more sophisticated calculations employing a multiple scattering description for electron tunneling \cite{palotas05}.
These peaks originate from minority $d$ electrons, in agreement with Refs.\ \cite{hofer00Fe,ferriani10tip}.
Left and right parts of Figure \ref{Fig2} show spectra obtained with assumed tip magnetization direction
antiparallel ($C=cos\varphi=-1$) and parallel ($C=cos\varphi=+1$) to that of Fe, respectively.
Note that the total and topographic tunneling spectra are always positive in the whole energy range,
while the magnetic contribution can also be negative.
Moreover, the topographic contribution equals in both parts as it is independent of tip magnetization, 
whereas the magnetic contribution changes sign, and the total $dI/dV$ curve changes accordingly.
While it is enhanced and its peaks are more pronounced in the antiparallel setup, it is lowered and
considerably flattened in the parallel case, the peak at $+0.2$ V even disappears.
This is an evidence that by modifying the magnetization direction of tip, the surface state peak can be tuned, and generally
suggests that the sensitivity of STS on magnetic samples can be enhanced by choosing tip magnetization direction properly.
For achieving this goal, and due to Figure \ref{Fig2}, we identify the magnetic contribution, $dI_{MAGN}/dV$ to play the key role.

In order to understand this mechanism more, we study electronic properties of the surface Fe atom,
$P_S(E),(n_S(E)+m_S(E)cos\varphi),n_S(E),m_S(E)cos\varphi$ (dashed lines in Figure \ref{Fig2}).
The calculated spin-polarization is $-0.79$ at the Fermi level, i.e.\ it is negative as has been reported recently by Ferriani
et al.\ \cite{ferriani10tip}, where the full-potential linearized augmented plane wave method has been applied.
The shape of our energy dependent spin-polarization is also in good agreement with Ferriani's result.
Comparing the remaining listed electronic properties to tunneling spectra it is clearly seen that they are highly correlated,
i.e.\ $dI_{TOTAL}/dV\propto (n_S+m_Scos\varphi)$, $dI_{TOPO}/dV\propto n_S$, $dI_{MAGN}/dV\propto m_Scos\varphi$,
each pair drawn with the same color in Figure \ref{Fig2}.
The multiplicative factor connecting each of these pairs is the energy dependent transmission coefficient
for tunneling electrons, $e^{-2\kappa(E)z}$, where we applied Eq.(\ref{Eq_kappa_TH}) for the energy dependent vacuum decay.
In Figure \ref{Fig2} the corresponding electronic properties are shown multiplied by $e^{-2\kappa(E_F^S)z}$, and rescaled
in the same way as $dI/dV$, such that e.g.
\begin{equation}
\label{Eq_prop}
\frac{dI_{MAGN}}{dV}(U=0)=e^{-2\kappa(E_F^S)z}m_S\left(E_F^S\right)cos\varphi=e^{-\frac{2z}{\hbar}\sqrt{2m\phi_S}}m_S\left(E_F^S\right)cos\varphi,
\end{equation}
and similarly for the rest of the correlated pairs.
Note that the conversion between bias voltage and energy is $eU=E-E_F^S$, i.e.\ $U=0$ corresponds to $E_F^S$.
Since the considered vacuum decay, Eq.(\ref{Eq_kappa_TH}), is monotonously decreasing with increasing energy, the relation
between the rescaled electronic properties and the corresponding $dI/dV$ spectra is the following: For positive bias voltages,
\begin{eqnarray}
\frac{dI_{TOTAL}}{dV}(U>0)&>&e^{-2\kappa\left(E_F^S\right)z}\left[n_S\left(E_F^S+eU\right)+m_S\left(E_F^S+eU\right)cos\varphi\right],\nonumber\\
\frac{dI_{TOPO}}{dV}(U>0)&>&e^{-2\kappa\left(E_F^S\right)z}n_S\left(E_F^S+eU\right),\nonumber\\
\left|\frac{dI_{MAGN}}{dV}(U>0)\right|&>&e^{-2\kappa\left(E_F^S\right)z}\left|m_S\left(E_F^S+eU\right)cos\varphi\right|,
\label{Eq_dIdV_rel_posV}
\end{eqnarray}
since $e^{-2\kappa(E_F^S+eU)z}>e^{-2\kappa(E_F^S)z}$. Here, we employed Eqs.(\ref{Eq_dIdV-idealtip}) and (\ref{Eq_dIdV-V}).
Similarly, for negative bias voltages, $e^{-2\kappa(E_F^S+eU)z}<e^{-2\kappa(E_F^S)z}$, and, thus,
\begin{eqnarray}
\frac{dI_{TOTAL}}{dV}(U<0)&<&e^{-2\kappa\left(E_F^S\right)z}\left[n_S\left(E_F^S+eU\right)+m_S\left(E_F^S+eU\right)cos\varphi\right],\nonumber\\
\frac{dI_{TOPO}}{dV}(U<0)&<&e^{-2\kappa\left(E_F^S\right)z}n_S\left(E_F^S+eU\right),\nonumber\\
\left|\frac{dI_{MAGN}}{dV}(U<0)\right|&<&e^{-2\kappa\left(E_F^S\right)z}\left|m_S\left(E_F^S+eU\right)cos\varphi\right|,
\label{Eq_dIdV_rel_negV}
\end{eqnarray}
This bias dependent relation between the correlated pairs can clearly be seen in Figure \ref{Fig2} (compare solid and dashed lines
of the same color), which is, in turn, due to the energy dependent vacuum decay.

From Figure \ref{Fig2} we identified the magnetic contribution, $dI_{MAGN}/dV$ to be responsible for tuning the STS peaks.
At a given energy $E$ this is, in effect, the tunneling transmission coefficient ($e^{-2\kappa(E)z}$) times the MDOS
of surface Fe ($n_S(E)P_S(E)$) times the cosine of the angle between spin quantization axes of surface and tip ($cos\varphi$).
Since $e^{-2\kappa(E)z}$ and $n_S(E)$ are always positive, the sign of $dI_{MAGN}/dV$ is determined by $P_S(E)cos\varphi$.
Let us focus on the energy region $E_F^S-0.4eV<E<E_F^S+1.0eV$, where the spin-polarization is negative.
If tip magnetization is antiparallel (parallel) to that of Fe, $cos\varphi=-1$ ($cos\varphi=+1$), then $P_S(E)cos\varphi$
is positive (negative), and so is $dI_{MAGN}/dV$. Adding this term to $dI_{TOPO}/dV$, which is not affected by tip magnetization,
the total differential tunneling spectrum is enhanced if tip magnetization is antiparallel, and decreased if it is parallel to Fe.
This explains the main message of Figure \ref{Fig2}. Although the surface state peak is already well obtained by using a
nonmagnetic tip ($dI_{TOPO}/dV$), the $dI/dV$ signal can be further improved by setting the tip magnetization direction
antiparallel to that of Fe. Note that in our ideal tip the energy independent spin-polarization was set to $P_T=1$, thus,
the largest effect on $dI/dV$ occurs in this case. Similar, but reduced effect is expected if we set the tip spin-polarization to
$0<P_T<1$, whereas $dI/dV$ would be enhanced in the studied energy regime by applying a tip magnetization direction parallel to Fe
if $P_T<0$. Setting $P_T=0$ corresponds to a nonmagnetic tip, which results in $dI/dV$=$dI_{TOPO}/dV$.

In Figure \ref{Fig2} we show simulated tunneling spectra considering contribution from one surface Fe atom only.
However, in Eq.(\ref{Eq_dIdV-idealtip}) the summation over $\alpha$ should, in principle, be carried out over all surface atoms.
Since the tunneling probability decays exponentially with increasing tip-sample distance, it is expected that
a finite number of surface atoms should be enough to be included in the summation in order to obtain converged $dI/dV$ functions.
In Figure \ref{Fig3} we study this convergence of the simulated differential tunneling spectrum above an Fe atom 
in the bias range from $-1.0$ V to $+1.0$ V using the ideal tip.
We consider the case of antiparallel tip magnetization direction only,
and the spectra are rescaled in the same way as in Figure \ref{Fig2}.
We show $dI/dV$ spectra by including different number of surface Fe atoms in the summation over $\alpha$ in
Eq.(\ref{Eq_dIdV-idealtip}), i.e.\ one Fe atom in a $1\times 1$ surface unit cell (1Fe, dotted line),
nine Fe atoms in a $3\times 3$ surface cell (9Fe, dashed line), and twenty-five Fe atoms in a $5\times 5$ surface cell
(25Fe, solid line), where the spectrum is calculated above the central Fe atom with lateral coordinates $(x_0,y_0)$ in each case.
Since the magnetic surface unit cell is identical to the chemical unit cell, all surface Fe atoms have the same
local electronic structure ($PDOS$).
The spectrum obtained by one Fe contribution is the same as the one in the left part of Figure \ref{Fig2}
drawn by black solid line.
It is clearly seen that by including more atoms in the summation, the spectrum is growing with no change of the peak positions.
We find that convergence is rapid, i.e.\ calculating $dI/dV$ from a $5\times 5$ surface cell is sufficiently converged, and
contribution from all Fe atoms in a $7\times 7$ cell means a relative increment of less than $10^{-4}$ compared to the $5\times 5$
cell in the studied bias range.
One should keep in mind that the independent orbital approximation for vacuum decay of electron states is employed, and taking
into account orbital variations \cite{chen90} would alter the fine structure of our calculated spectra without changing the peak
positions.

We believe that within the model of the atom-superposition approach our finding to obtain convergence of the $dI/dV$ spectrum
with respect to the spatial extension of the sample surface contributions is generally valid for any sample surface.
The reason is the exponential factor describing electron tunneling transmission, which is decaying rapidly as the
tip apex-surface atom distance increases. Considering heterogeneous sample surfaces, we always take into account the full chemical
unit cell closest to the tip apex position for summation over $\alpha$, and, thus, all important peaks appear in the $dI/dV$
spectrum. Convergence of $dI/dV$ with respect to spatial extension is obtained by involving atoms of neighboring full chemical
unit cells in the summation over $\alpha$.
Moreover, note that in our Fe(001) surface by including more atoms in the summation does not change our findings for the effect
of tip magnetization direction on the tunneling spectra described in Figure \ref{Fig2}.

In the above discussion we omitted energy variations of tip electronic structure for the purpose of studying only the effect of
tip magnetization direction on the differential tunneling spectrum. However, considering realistic tips the situation is somewhat
more complicated as $dI_{MAGN}/dV$ at a given energy $E$ is proportional to $n_T(E)P_T(E)n_S(E)P_S(E)cos\varphi$.
Let us study tunneling spectrum of the same Fe(001) surface by probing it with a ferromagnetic Ni tip.
Such tips are routinely used in SP-STM and SP-STS experiments \cite{heinrich09,heinrich10}.

The Ni tip has been modeled by a 7-layer Ni film slab with (110) orientation, having one Ni apex atom on both surfaces,
i.e.\ with a double vacuum boundary.
According to previous findings it is sufficient to assume one tip apex atom on top of a metal surface as tip model, and there is
no need to simulate more complex geometries, since the electronic structure of the tip apex does not change considerably, as it
has been shown e.g.\ for an Fe tip \cite{ferriani10tip}. For our purpose of demonstrating the effect of tip apex electronic
structure on SP-STS spectra, such simple tip model is sufficient.
In our tip the apex atom and the topmost surface layers have been relaxed on both sides. The interaction
between apex atoms in neighboring supercells is minimized by choosing a $3\times 3$ surface cell, and
a 15.4 $\AA$ wide separating vacuum region in $z$ direction.
Moreover, a $5\times 5\times 1$ MP k-point grid has been chosen for obtaining the projected DOS onto the apex atom.
Employing Eq.(\ref{Eq_WF_local}), the local electron workfunction above the tip apex is $\phi_T=4.52$ $eV$,
and Eq.(\ref{Eq_kappa_WKB}) has been used to determine the vacuum decay.

\begin{figure*}
\includegraphics[width=0.5\textwidth,angle=0]{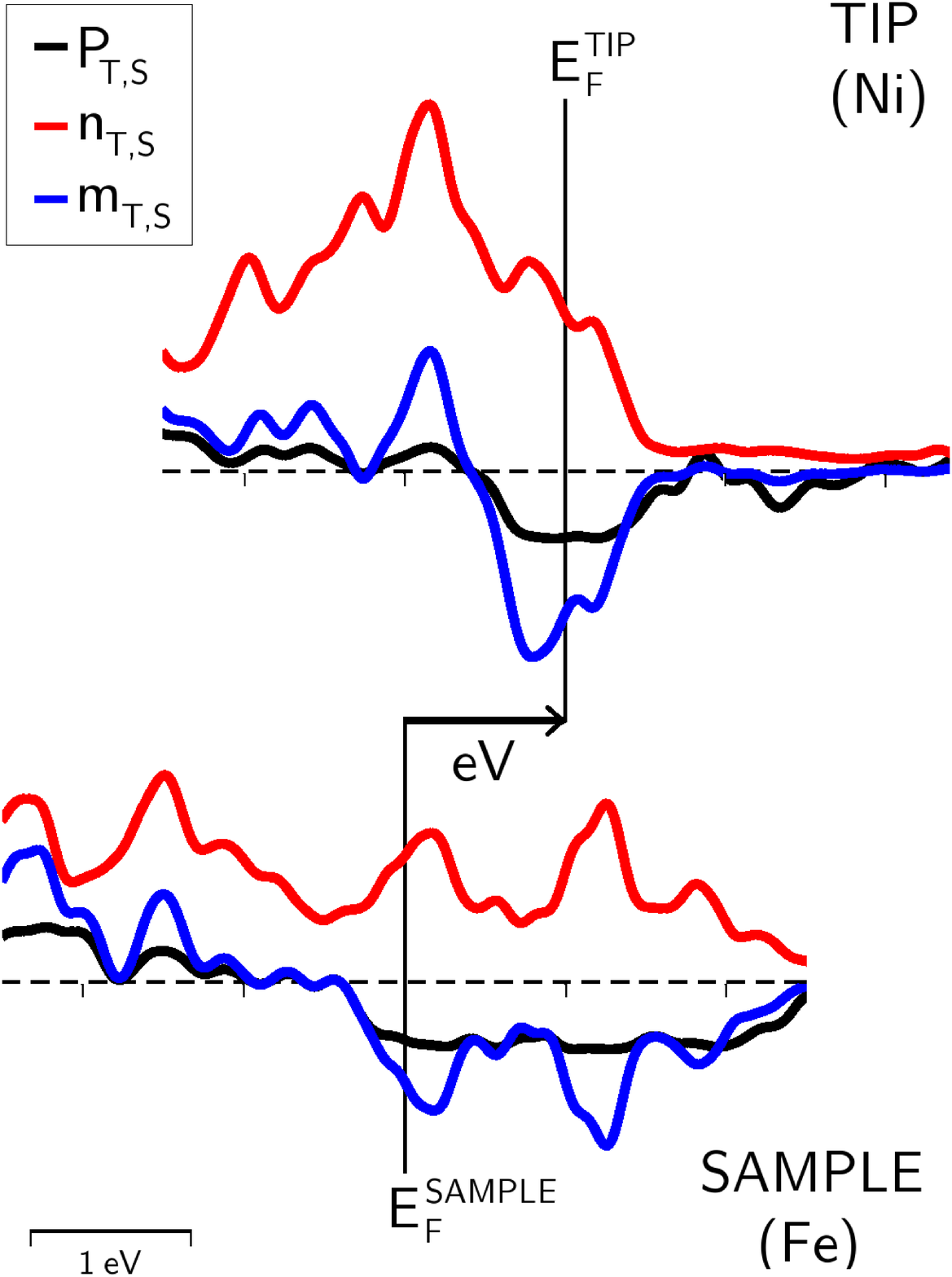}
\caption{\label{Fig1} (Color online) Sketch of the relative energetic position of sample and tip electronic structures.
$P(E),n(E),m(E)$ of the Ni tip apex and the surface Fe atom, and their relation at applied bias voltage, $V$ are shown.
Depending on $V$, different electron states are involved in the tunneling process
for calculating the tunneling current, while $V$ (bias set point) fixes the relative energetic position of tip and sample
electron states when simulating differential tunneling spectra, i.e.\ $E_F^{TIP}=E_F^{SAMPLE}+eV$. The
energy scale is shown in the bottom left part (1 electronvolt), and according to that the bias voltage in the Figure is +1.0 V.
}
\end{figure*}

\begin{figure*}
\includegraphics[width=1.0\textwidth,angle=0]{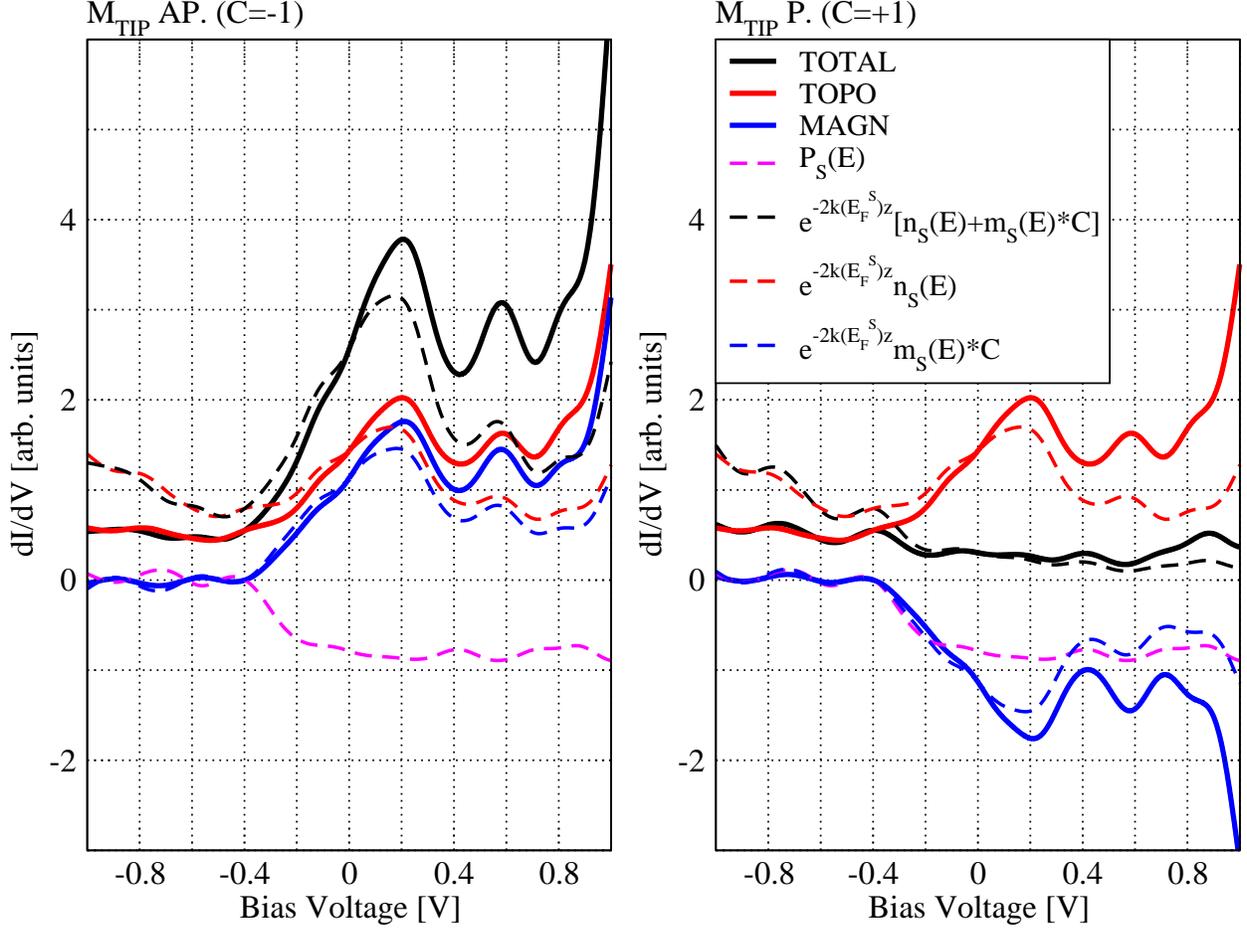}
\caption{\label{Fig2} (Color online) Simulated differential tunneling spectra (TOTAL) and their topographic (TOPO) and
magnetic (MAGN) contributions (solid lines) $z$=3.5 $\AA$ above an Fe atom on the Fe(001) surface, assuming a maximally
spin-polarized and electronically flat tip using Eq.(\ref{Eq_dIdV-idealtip}) and Eq.(\ref{Eq_dIdV_decomp}).
Tip magnetization direction is antiparallel (C=$cos\varphi=-1$) (left part) and parallel (C=$cos\varphi=+1$) (right part)
to that of Fe. Note that in the two parts the topographic contribution is the same, while the magnetic contribution changes sign,
and the total $dI/dV$ curve changes accordingly.
By modifying the magnetization direction of tip, the surface state peak can be tuned.
For comparison, spin-polarization, $P_S(E)$, and other electronic properties of the surface Fe atom,
$(n_S(E)+m_S(E)*cos\varphi),n_S(E),m_S(E)*cos\varphi$, each multiplied by $e^{-2\kappa(E_F^S)z}$ are shown in both parts
(dashed lines). These properties correlate well with the corresponding spectra (same color), see text for details.
}
\end{figure*}

\begin{figure*}
\includegraphics[width=1.0\textwidth,angle=0]{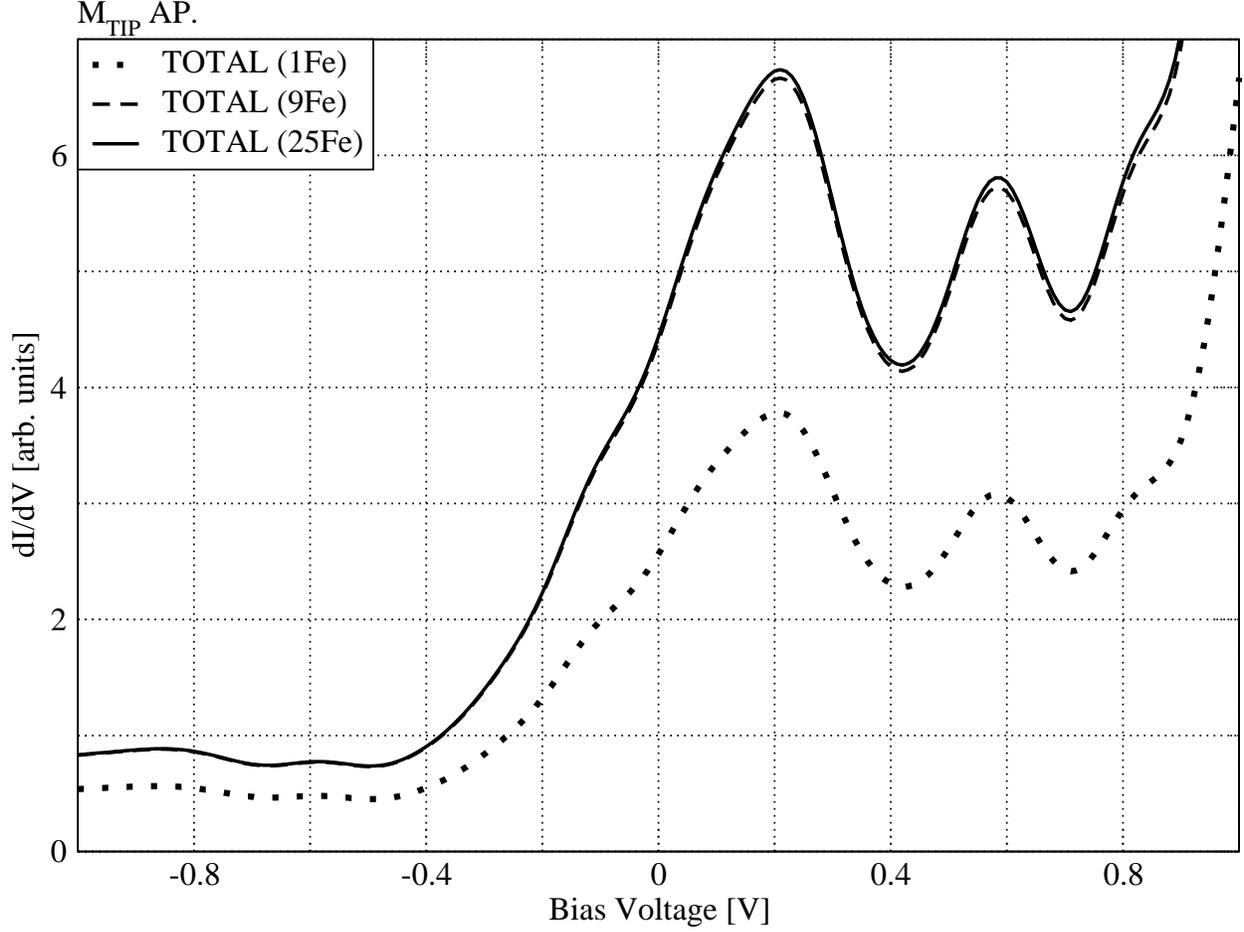}
\caption{\label{Fig3} Simulated differential tunneling spectra 3.5 $\AA$ above an Fe atom on the Fe(001) surface, and its
convergence depending on the contribution from the number of Fe atoms in the summation over $\alpha$ in
Eq.(\ref{Eq_dIdV-idealtip}), i.e.\ from one Fe atom (1Fe, dotted line) in a $1\times 1$ surface unit cell (the spectrum is taken
above this atom), from nine Fe atoms (9Fe, dashed line) in a $3\times 3$ surface cell (the spectrum is above the central Fe),
and from twenty-five Fe atoms (25Fe, solid line) in a $5\times 5$ surface cell (the spectrum is above the central Fe).
Note that each surface Fe atom has the same local electronic structure.
For the simulation a maximally spin-polarized and electronically flat tip is applied using Eq.(\ref{Eq_dIdV-idealtip}), and
tip magnetization direction is antiparallel to that of Fe.
}
\end{figure*}

Figure \ref{Fig1} shows the calculated electronic structure, $P(E),n(E),m(E)$ of the Ni tip apex atom (top part)
and the Fe surface atom (bottom part) in the energy range $[-2.5eV,+2.5eV]$ with respect to the corresponding Fermi energies.
For Fe we find additional important $PDOS$ peaks outside the $[E_F^S-1eV,E_F^S+1eV]$ energy range reported in Figure \ref{Fig2}.
Peaks at $E_F^S+1.25eV$ and $E_F^S+1.8eV$ have minority $d$ character, while the peak at $E_F^S-1.5eV$ originates mostly from
majority $d$ electrons, all these in good agreement with Refs.\ \cite{hofer00Fe,ferriani10tip}.
Focusing on Ni, we find that $n_T(E)$ is almost constant above $E_F^{TIP}+0.6eV$, and most importantly, the spin-polarization is
$-0.91$ at the Fermi level, $E_F^{TIP}$, and it is negative and high in absolute value, i.e.\ $|P_T(E)|>0.8$ between
$E_F^{TIP}-0.3eV$ and $E_F^{TIP}+0.3eV$.
The energetic relation of tip and sample electronic structures is determined by the bias voltage, $V$, illustrated in
Figure \ref{Fig1}, where $V$=+1.0 V has been chosen.
As $E_F^{TIP}=E_F^S+eV$, the whole tip electronic structure is shifted by $eV$ with respect to that of the sample.
This means that depending on $V$, different electron states are involved in the tunneling process
for calculating the tunneling current. Simulating differential tunneling spectra, the bias voltage is called bias set point and
it fixes the relative energetic position of tip and sample electron states.
As we learned from Figure \ref{Fig2}, the total $dI/dV$ signal can be tuned depending on the magnetic contribution,
$dI_{MAGN}/dV$, which is proportional to $n_T(E)P_T(E)n_S(E)P_S(E)cos\varphi$ at a given energy $E$, in our model
including energy variations of tip electronic structure.
Depending on the bias set point the latter product can vary considerably, even can change sign, which, in effect, determines
whether $dI/dV$ is enhanced or decreased at the given energy. Since $n_T(E)$ and $n_S(E)$ are always positive, the
decisive factor for the sign of the magnetic contribution is the effective spin-polarization, $P_T(E)P_S(E)cos\varphi$.

\begin{figure*}
\includegraphics[width=1.0\textwidth,angle=0]{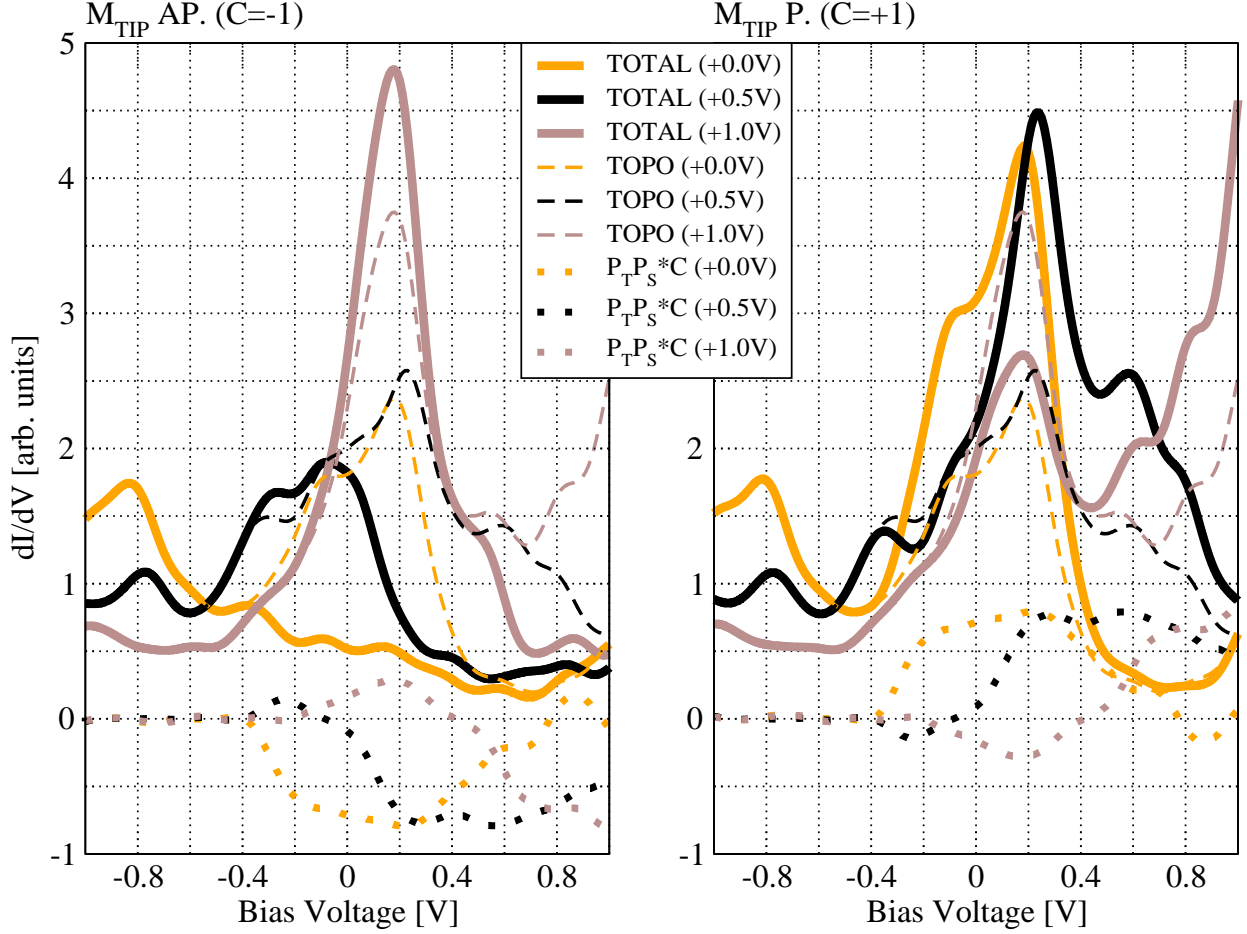}
\caption{\label{Fig4} (Color online) Simulated differential tunneling spectra (TOTAL, solid lines) and topographic contribution
(TOPO, dashed lines) 3.5 $\AA$ above an Fe atom on the Fe(001) surface, using a model Ni tip (see text for details), and
depending on the bias set point for the tip (curves with different colors). The chosen bias set points are given in parentheses.
They fix the relative energetic position of tip and sample electron states, see Figure \ref{Fig3}.
Tip magnetization direction is antiparallel (C=$cos\varphi=-1$) (left part) and parallel (C=$cos\varphi=+1$) (right part)
to that of Fe.
Note that the topographic contributions for the corresponding bias set point are the same for both tip magnetization directions,
while the total $dI/dV$ curves change considerably, similarly to Figure \ref{Fig1}.
This is due to $P_T(E)P_S(E)cos\varphi$ variations depending on the bias set point, which are shown with dotted lines.
Thus, by modifying the bias set point, the surface state peak can be tuned.
}
\end{figure*}

In order to illustrate this effect, we consider three different bias set points taking
contribution from one surface Fe atom to $dI/dV$ in Figure \ref{Fig4}.
This Figure shows simulated differential tunneling spectra (TOTAL, solid lines) and topographic contribution
(TOPO, dashed lines) above the surface Fe atom in the bias range from $-1.0$ V to $+1.0$ V using our model Ni tip,
employing Eq.(\ref{Eq_dIdV}), depending on the bias set point, $0.0$ V, $+0.5$ V, $+1.0$ V, indicated by different colors.
The spectra are rescaled such that they can be shown together with the effective spin-polarization, $P_T(E)P_S(E)cos\varphi$,
drawn by dotted lines.
The topographic contributions for the corresponding bias set point are the same for both tip magnetization directions,
and the iron surface state peak position is more or less reserved, while the peak heights change depending on the bias set point.
This effect is due solely to bias set point, and not on tip magnetization, as $dI_{TOPO}/dV$ is proportional to $n_T(E)n_S(E)$
at energy $E$, and the bias set point determines the relative position of tip and sample electronic structures.
On the other hand, the total $dI/dV$ curves change considerably depending not only on bias set point but also on tip magnetization
direction, similarly to the observation assuming an electronically flat tip electronic structure, see Figure \ref{Fig2}.
Focusing on the antiparallel tip magnetization direction ($C=cos\varphi=-1$, left part of Figure \ref{Fig4}),
it is seen that the iron surface state peak disappears at $0.0$ V bias set point (orange solid line),
it is shifted to $-0.10$ V by using $+0.5$ V (black solid line),
while it remains at the same position and is enhanced by using $+1.0$ V (brown solid line).
On the other hand, the effect is different by setting tip magnetization direction parallel to that of Fe
($C=cos\varphi=+1$, right part of Figure \ref{Fig4}). Here, for all considered bias set points the peak position remains, and for
$0.0$ V and $+0.5$ V the peak height is enhanced, while for $+1.0$ V the peak height is decreased.
All these findings show an evidence that the STS peaks can be tuned by changing the bias set point. 
The reason for this effect is $n_T(E)n_S(E)$ for the topographic part, and $P_T(E)P_S(E)cos\varphi$ for the
magnetic part, whose sign determines the sign of $dI_{MAGN}/dV$ and, thus, the total $dI/dV$ related to $dI_{TOPO}/dV$.
This is clearly seen for all considered cases: if $P_T(E)P_S(E)cos\varphi>0$ then $dI/dV>dI_{TOPO}/dV$ in the corresponding energy
regime, and similarly, if $P_T(E)P_S(E)cos\varphi<0$ then $dI/dV<dI_{TOPO}/dV$.

The results suggest that the sensitivity of STS on a magnetic sample can be enhanced by choosing the appropriate bias set point
and magnetic tip. $dI_{TOPO}/dV$ peaks are expected at local maxima of $n_T(E)n_S(E)$, while the maximal $dI_{MAGN}/dV$ can
be obtained in case of having parallel tip and sample spin-polarization vectors with $+1$ value of the spin-polarization, each.
This effect should also be observed in STS experiments. By designing the proper tip material,
possibly with the help of electronic structure calculations, i.e.\ according to our study, or e.g.\ Ref.\ \cite{ferriani10tip},
the advantage on the sensitivity of SP-STS measurements can be expected.
However, the controlled preparation of magnetic tips for SP-STS experiments is not at all an easy task \cite{rodary11}.
Since our model assumes one tip apex atom, at present, we can not take rough tip structures or nanotips into account.
On the other hand, it is naturally possible to simulate such rough tip structures within our atom-superposition framework
by considering summation over different tip atoms contributing to $dI/dV$. This could be a research direction in the future.
Though our present study does not help for improving tip preparation techniques, it is demonstrated that the governing factor
of the fine structure and sensitivity of SP-STS spectra is the effective spin-polarization.
Note that above considerations are generally valid for any combination of magnetic sample and tip, and could explain the
observed difference in the bias dependent structure of the measured STS spectra above Co islands of opposite magnetization
\cite{schouteden08,heinrich10}.

In the present study we omitted energy variations of tip and sample spin quantization axes. However, in certain combinations of
tip and sample the situation is even more complicated as $dI_{MAGN}/dV$ at a given energy $E$ is proportional to
$n_T(E)P_T(E)n_S(E)P_S(E)cos\varphi(E)$, i.e.\ the angle between spin quantization axes can also depend on energy.
We will address this question in the future.

\section{Conclusions}
\label{sec_conc}

Motivated by recent SP-STS experiments we studied tip effects on the tunneling spectrum of a model magnetic Fe(001) surface.
By considering an ideal electronically flat and maximally spin-polarized tip we found that STS peaks are sensitive to the
tip magnetization direction relative to the surface. In case of a model Ni tip the role of the bias set point for the tip is
highlighted, which fixes the relative energetic position of sample and tip electronic structures.
We showed evidence that the fine structure of the tunneling spectrum is governed by the effective spin-polarization.
In conclusion, our results suggest that the sensitivity of STS on a magnetic sample can be tuned and even enhanced by choosing
the combination of appropriate magnetic tip and bias set point.

\section{Acknowledgments}

Financial support of the Magyary Foundation, EEA and Norway Grants, the Hungarian Scientific Research Fund (OTKA PD83353, K77771)
and the New Hungary Development Plan (Project ID: T\'AMOP-4.2.1/B-09/1/KMR-2010-0002) is gratefully acknowledged.

\newpage

\end{document}